\documentclass[reprint,noshowpacs,noshowkeys,prd,balancelastpage,nofootinbib]{revtex4}
\usepackage{amsfonts}
\usepackage{mdframed}
\usepackage{amssymb}
\usepackage{footnote}
\usepackage{amsmath}
\usepackage{graphicx}
\usepackage{float}
\usepackage[font={footnotesize,it}]{caption}
\usepackage[utf8]{inputenc}
\usepackage{natbib}
\usepackage{tikz}
\usepackage{tikz-3dplot}
\usepackage[colorlinks=true,
            linkcolor=purple,
            urlcolor=purple,
            citecolor=blue]{hyperref}

\begin{document}

\title{Generic spherically symmetric thin-shell wormholes with equilibrium
throat at the innermost photonsphere are unstable}
\author{S. Habib Mazharimousavi}
\email{habib.mazhari@emu.edu.tr}
\affiliation{Department of Physics, Faculty of Arts and Sciences, Eastern Mediterranean
University, Famagusta, North Cyprus via Mersin 10, T\"{u}rkiye}

\begin{abstract}
In this research, we prove analytically that a generic spherically symmetric
thin-shell wormhole (TSW) with its throat located at the innermost
photonsphere of the bulk asymptotically flat black hole and supported by a
generic surface barotropic perfect fluid is unstable against a radial linear
perturbation. This is the generalization of the instability of the
Schwarzschild TSW (STSW) with the throat's radius located at $a_{0}=3M$ that
was revealed by Poisson and Visser in their seminal work \cite{Poisson1995}
where they studied the mechanical stability of STSW. Our proof provides a
link between the instability of the null circular geodesics on the innermost
photonsphere of a generic static spherically symmetric asymptotically black
hole and the TSW constructed in the same bulk with $a_{0}=r_{c}$ where $%
a_{0} $ and $r_{c}$ are the radius of the TSW and the innermost
photonsphere, respectively. For asymptotically flat spherically symmetric
black holes possessing more than one photonspheres, the number of the
photonspheres is odd and at least one photonsphere is stable which implies
the corresponding TSW with its throat identical with the stable photonsphere
is also stable.
\end{abstract}

\date{\today }
\maketitle

\section{Introduction}

Wormholes have been known to be solutions of Einstein's equations from the
early of General Relativity (GR) in the works of Flamm in \cite{Flamm1916}
and Einstein and Rosenin \cite{Einstein1935} that is also called the
Einstein-Rosen Bridge (ERB). It became popular after the seminal works of
Morris and Thorne in \cite{Morris1988,Thorne1988}. The newly popularized
wormholes were also the direct solutions of Einstein's field equations,
however, unlike the former one they were traversable \cite%
{Morris1988,Thorne1988}. These traversable wormholes in the standard GR are
supported by an exotic energy-momentum tensor \cite%
{Morris1988,Thorne1988,Visser1996,Visser1998}. Such exotic matters do not
satisfy the null energy condition (NEC) which is known to be respected by
all types of known normal or physical matters. In recent developments, the
requirement of an exotic matter in traversable wormhole spacetimes has been
removed in some specific modified theories of gravity. For more details, we
refer to \cite{N1,N2,N3,N4,N5,N6,N7,N8,N9,N10,N11}.

Apart from the traditional wormhole which is directly the solution of
Einstein's field equations, there is another type of wormhole that is not a
direct solution to Einstein's field equations. Such wormholes have been
introduced by Matt Visser in \cite{MV1,MV2}, using the Israel junction
formalism \cite{Israel1967}. The throat of this new traversable wormhole is
a thin shell connecting the two geodesically incomplete spacetimes which are
cut on a timelike hypersurface from an identical or two individual solutions
of Einstein's equations. At the timelike boundary hypersurface where the two
spacetimes are joined, there exists an exotic surface fluid such that the
throat is effectively a physical thin shell instead of a mathematical
concept. Therefore such wormholes are known as thin-shell wormholes (TSWs).
The fundamental difference between the wormholes and TSW is that while
wormholes are supported by exotic matter that fills the entire spacetime,
TSWs are powered by exotic matters that are accumulated at only their
throat, and off the throat, everything is as normal as the corresponding
bulk spacetime. This should be added that, in analogy with the standard
wormholes, here for TSW the surface fluid present on the throat may also be
physical in some modified theories of gravity \cite{TN1,TN2,TN3}.

Conceptually, one of the reasons that makes TSWs important is their similar
physical structure to the bulk spacetime where they are constructed. For
instance, the Schwarzschild TSW \cite{Poisson1995} possesses common features
with the Schwarzschild black hole. Particularly, the null or timelike
geodesics of the two, in most of the cases, are the same which makes it
difficult to distinguish one from the other. To be more specific, we refer
to the paper of Diemer and Smolarek in \cite{Diemer2013} where the dynamics
of test particles in the STSW as well as the rotating Kerr TSW have been
investigated. It can be seen in \cite{Diemer2013} that the geodesics of a
null or a timelike particle that doesn't cross the throat and remains on one
side of the throat is the same as the geodesics of an identical particle
with identical initial conditions in the Schwarzschild black hole.
Furthermore, those particles that due to their initial conditions cross the
throat and transmit to the other side of the throat (other universe) make a
symmetric geodesics with respect to the throat (see Fig. 2 and Fig. 3 in 
\cite{Diemer2013} ). The geodesics on each side of the throat are also
identical to the geodesics of an identical particle in the Schwarzschild
black hole, up to the radial distance from the horizon coinciding with the
location of the throat in the STSW. Such particles, after this point,
continue their geodesics toward the horizon and beyond. Now suppose that the
throat of the STSW is very close to the event horizon of the Schwarzschild
black hole. In such a case, from a distant observer's frame, the two
geodesics are not easily distinguishable. Therefore, hypothetically one may
consider a static black hole to be a TSW whose event horizon is replaced by
a throat and the falling particles move to the other universe instead of the
central singularity. This could also solve the problem of the singularity of
the black holes. Let's add that in the latter case where the throat
approaches the event horizon the corresponding TSW may not be considered
traversable anymore as the tidal force near the throat becomes unbearable
for a traveler.

Furthermore, similar to the black hole spacetimes, the mechanical stability
of TSWs has also attracted attention from the beginning of its introduction 
\cite{S1,S2,S3,S4,S5,S6,S7,S8,S9,S10,S11,S12,S13,S14,S15,S16}. Varela in 
\cite{Varela2015} studied the instability of the Schwarzschild thin-shell
wormhole (STSW) with the radius of the throat at $a_{0}=3M$ which was
introduced by Poisson and Visser in \cite{Poisson1995}. This instability was
known from \cite{Poisson1995} where the STSW was supported by a surface
fluid with a generic barotropic Equation of State (EoS). In \cite{Varela2015}
by virtue of the concept of variable EoS the STSW with $a_{0}=3M$ became
stable. Let us add that while the generic barotropic EoS implies $p=p\left(
\sigma \right) ,$ the variable EoS (see \cite{Lobo2012}) is described by $%
p=p\left( \sigma ,a\right) $ where $\sigma ,$ $p,$ and $a$ are the
surface-energy density, the surface transverse pressure, and the radius of
the throat in a generic TSW, respectively.

It is well known that for the Schwarzschild black hole the radius of the
innermost circular null geodesics is $r_{c}=3M$ while its event horizon is
located at $r=r_{s}=2M.$ Referring to the original work of Poisson and
Visser \cite{Poisson1995}, we observe that\ the mechanical stability of the
STSW is divided into two distinct regions at $a_{0}=3M$ (see Eqs. (31) and
(32) as well as Fig. 1 of Ref. \cite{Poisson1995}). This seems not to be a
coincidence that the definite instability radius of the STSW and the
innermost photonsphere are both at $a_{0}=3M$ and $r_{c}=3M.$ \textit{Is
there a deeper connection between the innermost photonsphere of generic
spherically symmetric black holes and the instability of the corresponding
TSW}? Answering this question is our main motivation for performing this
current research. In the next Section, first of all, we show that for a
generic asymptotically flat static spherically symmetric black hole there
always exists a circular null geodesics that is called the innermost light
ring. In \cite{D1} the existence of such a circular orbit in arbitrary
dimensions has been proved. It was also shown that such an innermost null
circular ring is unstable \cite{D1}. More recently Peng in \cite{P1} has
also proved the existence of such photon ring for the extremal spherically
symmetric asymptotically flat hairy black holes. In a different approach,
Hod in \cite{H3} has also proved the same statement while for non-extremal
cases he proved the existence of circular orbit in \cite{H0} (see also \cite%
{P2}). The null geodesics and light rings outside static black holes are the
indirect sources of information about the corresponding black holes. This is
because of the strong gravity of the black holes such that even light cannot
escape from them. Therefore direct effects of black holes cannot be
observed. Some of the important works on the indirect effects of black holes
can be found in \cite{D1,P1,H3,H0,P2,H1}\ (also see \cite%
{7,8,9,10,11,12,13,14,15,16,17,18,19,20} and the references therein).

\section{Generic Spherical Thin-Shell Wormhole with its throat at the
photonsphere of the bulk spacetime}

Inspired by the great works in \cite{D1,P1,H3,H0,P2,H1}, we start with a
generic static spherically symmetric bulk spacetime described by%
\begin{equation}
ds^{2}=-\psi \left( r\right) dt^{2}+\frac{dr^{2}}{\psi \left( r\right) }%
+r^{2}\left( d\theta ^{2}+\sin ^{2}\theta d\varphi ^{2}\right) ,  \label{1}
\end{equation}%
in which $\psi \left( r_{+}\right) =0$ introduces the possible event horizon
located at $r=r_{+}.$ Let us add that (\ref{1}) is not the most generic
spherically symmetric line element as we imposed $g_{tt}g_{rr}=-1$\ which
corresponds to a particular matter field with the energy-momentum tensor $%
T_{\mu }^{\nu }=diag\left[ -\rho ,p,q,q\right] $\ in which $\rho +p=0.$

The outer photonsphere of the metric (\ref{1}) is defined as the circular
geodesics of the null particles, namely photons. To find the location of
such photonsphere, we write down the Lagrangian of the null particle which
is given by 
\begin{equation}
2\mathcal{L}=-\psi \left( r\right) \dot{t}^{2}+\frac{\dot{r}^{2}}{\psi
\left( r\right) }+r^{2}\left( \dot{\theta}^{2}+\sin ^{2}\theta \dot{\varphi}%
^{2}\right) ,  \label{2}
\end{equation}%
where a dot implies a derivative with respect to an affine parameter. Having
the Lagrangian defined in (\ref{2}), we obtain the components of the
generalized momentum which are given as follows 
\begin{equation}
P_{t}=-E=\frac{\partial \mathcal{L}}{\partial \dot{t}}=-\psi \left( r\right) 
\dot{t},  \label{3}
\end{equation}%
\begin{equation}
P_{\varphi }=L=\frac{\partial \mathcal{L}}{\partial \dot{\varphi}}=r^{2}\sin
^{2}\theta \dot{\varphi},  \label{4}
\end{equation}%
\begin{equation}
P_{r}=\frac{\partial \mathcal{L}}{\partial \dot{r}}=\frac{\dot{r}}{\psi
\left( r\right) },  \label{5}
\end{equation}%
and%
\begin{equation}
P_{\theta }=\frac{\partial \mathcal{L}}{\partial \dot{\theta}}=r^{2}\dot{%
\theta}.  \label{6}
\end{equation}%
As it is seen from (\ref{2}), the Lagrangian is independent of $t$ and $%
\varphi $ implying that the energy $E$ and the angular momentum $L$ of the
null particle, are conserved. Following the definition of the Hamiltonian of
the system i.e., $\mathcal{H}=\Sigma P_{\mu }\dot{x}^{\mu }-\mathcal{L},$ we
obtain the Hamiltonian given by%
\begin{equation}
2\mathcal{H}=-E\dot{t}+\frac{\dot{r}^{2}}{\psi \left( r\right) }+r^{2}\dot{%
\theta}^{2}+L\dot{\varphi}.  \label{7}
\end{equation}%
Furthermore, for a null particle $\mathcal{H}=0$ which consequently yields%
\begin{equation}
\dot{r}^{2}=\psi \left( r\right) \left( E\dot{t}-r^{2}\dot{\theta}^{2}-L\dot{%
\varphi}\right) .  \label{8}
\end{equation}%
Next, we set $\theta =\frac{\pi }{2}$ that after introducing 
\begin{equation}
\dot{t}=\frac{E}{\psi \left( r\right) },  \label{9}
\end{equation}%
and%
\begin{equation}
\dot{\varphi}=\frac{L}{r^{2}},  \label{10}
\end{equation}%
in (\ref{8}) we obtain%
\begin{equation}
\dot{r}^{2}=\psi \left( \frac{E^{2}}{\psi }-\frac{L^{2}}{r^{2}}\right) .
\label{11}
\end{equation}%
Finally, for circular geodesics, we impose $\dot{r}^{2}=0$ as well as $\frac{%
d}{dr}\left( \dot{r}^{2}\right) =0.$ Hence, we end up with the equation%
\begin{equation}
2\psi =r\psi ^{\prime },  \label{12}
\end{equation}%
and the ratio of the energy over angular momentum is given by%
\begin{equation}
\left( \frac{E}{L}\right) ^{2}=\frac{\psi }{r^{2}}.  \label{13}
\end{equation}%
In (\ref{12}) a prime notation stands for the derivative with respect to $r$%
. We note that (\ref{12}) is not a differential equation but an ordinary
equation for identifying the radius of the photonsphere i.e., $r=r_{c}.$
Considering the bulk spacetime to be an asymptotically flat black hole, we
impose $\psi \left( r_{+}\right) =0$ and $\psi \left( r\rightarrow \infty
\right) \rightarrow 1$. Moreover, Hod in \cite{Hod2011} has proved that the
circular orbits on the equatorial plane are the orbits with locally the
shortest period (as measured by an asymptotic observer). Herein, the period
of a null particle on the circular motion measured by an asymptotic observer
is simply found by setting $\Delta s=\Delta r=\Delta \theta =0$ and $\Delta
\varphi =2\pi $ in (\ref{1}) which yields%
\begin{equation}
T=\left. \frac{2\pi r}{\sqrt{\psi }}\right\vert _{r=r_{c}}.  \label{14}
\end{equation}%
To find the minimum of $T,$ we apply%
\begin{equation}
\left. T^{\prime }\left( r\right) \right\vert _{r=r_{c}}=0,  \label{15}
\end{equation}%
which explicitly becomes%
\begin{equation}
\left. 2\psi =r\psi ^{\prime }\right\vert _{r=r_{c}},  \label{16}
\end{equation}%
that is the same equation as (\ref{12}). For Eq. (\ref{16}) there exists at
least one solution due to the behavior of the period of the circular orbit
in Eq. (\ref{14}) where we observe that $T\left( r_{c}\rightarrow
r_{+}\right) \rightarrow \infty $ and $T\left( r_{c}\rightarrow \infty
\right) \rightarrow \infty $. This implies that the period $T$ at least for
one critical $r_{c}$ becomes minimum which is the solution of the equation (%
\ref{16}) too. Having known that (\ref{16}) admits at least one solution, we
conclude that (\ref{12}) also admits at least one solution which is, in
fact, the radius of the innermost circular orbit. In summary, we have
briefly shown that the innermost photonsphere exists and its radius
satisfies Eqs. (\ref{12}) and (\ref{16}).

Next, we construct a thin-shell wormhole in the bulk spacetime (\ref{1}) by
applying the standard Israel junction formalism \cite{Israel1967}. Without
going through the details, we refer to the rich literature on the subject
and only use their results \cite%
{S1,S2,S3,S4,S5,S6,S7,S8,S9,S10,S11,S12,S13,S14,S15,S16}. Considering the
throat of the TSW to be at $r=a>r_{+}$ the induced line element of the
throat surface is given by%
\begin{equation}
ds^{2}=-d\tau ^{2}+a\left( \tau \right) ^{2}\left( d\theta ^{2}+\sin
^{2}\theta d\varphi ^{2}\right) ,  \label{17}
\end{equation}%
in which $\tau $ is the proper time on the throat. To join the spacetimes of
either side of the TSW at the throat smoothly, one has to apply the Israel
junction conditions \cite{Israel1967} upon which a surface energy-momentum
tensor is required. This surface energy-momentum tensor is given by $S_{\mu
}^{\nu }=diag\left( -\sigma ,p,p\right) $ in which the surface energy
density $\sigma $ and the surface transverse pressure $p$ respectively are
given by%
\begin{equation}
\sigma =-\frac{\sqrt{\psi \left( a\right) +\dot{a}^{2}}}{2\pi a},  \label{18}
\end{equation}%
and%
\begin{equation}
p=\frac{\sqrt{\psi \left( a\right) +\dot{a}^{2}}}{8\pi }\left( \frac{2\ddot{a%
}+\psi ^{\prime }\left( a\right) }{\psi \left( a\right) +\dot{a}^{2}}+\frac{2%
}{a}\right) ,  \label{19}
\end{equation}%
where a dot stands for a derivative with respect to the proper time $\tau .$
Moreover, the energy density $\sigma $ and the pressure $p$ satisfy the
energy conservation equation given by%
\begin{equation}
\frac{d\sigma }{da}+\frac{2}{a}\left( \sigma +p\right) =0.  \label{20}
\end{equation}%
Next, we assume that $a=r_{c}$ is an equilibrium radius for the throat such
that $\dot{a}=\ddot{a}=0.$ The equilibrium surface energy density and
transverse pressure are then found to be%
\begin{equation}
\sigma _{c}=-\frac{\sqrt{\psi \left( r_{c}\right) }}{2\pi r_{c}},  \label{21}
\end{equation}%
and%
\begin{equation}
p_{c}=\frac{\sqrt{\psi \left( r_{c}\right) }}{8\pi }\left( \frac{\psi
^{\prime }\left( r_{c}\right) }{\psi \left( r_{c}\right) }+\frac{2}{r_{c}}%
\right) .  \label{22}
\end{equation}%
Considering, $r_{c}$ the radius of the innermost photonsphere, Eq. (\ref{16}%
) implies 
\begin{equation}
\frac{\psi ^{\prime }\left( r_{c}\right) }{\psi \left( r_{c}\right) }=\frac{2%
}{r_{c}}  \label{R1}
\end{equation}%
and consequently 
\begin{equation}
p_{c}=-\sigma _{c}.  \label{23}
\end{equation}%
Now we study the stability of the TSW against a linear radial perturbation
about the equilibrium radius of the throat. To do so, we rewrite Eq. (\ref%
{18}) to get 
\begin{equation}
\dot{a}^{2}+V\left( a\right) =0,  \label{24}
\end{equation}%
in which the effective potential of the one-dimensional motion (\ref{24}) is
given by%
\begin{equation}
V\left( a\right) =\psi \left( a\right) -\left( 2\pi a\sigma \right) ^{2}.
\label{25}
\end{equation}%
For a linear perturbation, we expand the effective potential about its
equilibrium radius such that 
\begin{equation}
V\left( a\right) =V\left( r_{c}\right) +V^{\prime }\left( r_{c}\right)
\left( a-r_{c}\right) +\frac{1}{2}V^{\prime \prime }\left( r_{c}\right)
\left( a-r_{c}\right) ^{2}+\mathcal{O}\left( \left( a-r_{c}\right)
^{3}\right) ,  \label{26}
\end{equation}%
in which while $V\left( r_{c}\right) =\psi \left( r_{c}\right) -\left( 2\pi
r_{c}\sigma _{c}\right) ^{2}$ is zero, one calculates 
\begin{equation}
V^{\prime }\left( a\right) =\psi ^{\prime }\left( a\right) +8\pi ^{2}a\sigma
\left( \sigma +2p\right) .  \label{27}
\end{equation}%
where we used the energy-conservation equation (\ref{20}). Please note that
at the equilibrium where $a=r_{c}$, $\sigma _{c}$\ is given by (\ref{21})
which clearly yields from (\ref{25}) that $V\left( r_{c}\right) =0.$
Considering a generic barotropic EoS for the fluid present at the throat in
the form of $p=p\left( \sigma \right) $, we further calculate 
\begin{equation}
V^{\prime \prime }\left( a\right) =\psi ^{\prime \prime }\left( a\right)
-8\pi ^{2}\left\{ \left( \sigma +2p\right) ^{2}+2\sigma \left( \sigma
+p\right) \left( 1+2p^{\prime }\right) \right\} .  \label{28}
\end{equation}%
In (\ref{28}) we used the relation $\frac{dp\left( \sigma \right) }{da}=%
\frac{dp\left( \sigma \right) }{d\sigma }\frac{d\sigma }{da}$\ or
equivalently $\frac{dp\left( \sigma \right) }{da}=p^{\prime }\sigma ^{\prime
}=-\frac{2}{a}\left( \sigma +p\right) p^{\prime }$ where we used (\ref{20}).
At $a=r_{c}$, we impose $p_{c}=-\sigma _{c}$\ from (\ref{23}) such that 
\begin{equation}
V^{\prime }\left( r_{c}\right) =\psi ^{\prime }\left( r_{c}\right) -8\pi
^{2}r_{c}\sigma _{c}^{2}  \label{R2}
\end{equation}%
which after applying (\ref{R1}) and considering (\ref{21}) it becomes
identically zero. Furthermore from (\ref{28}) at $a=r_{c}$\ and $%
p_{c}=-\sigma _{c}$\ one obtains 
\begin{equation}
V^{\prime \prime }\left( r_{c}\right) =\psi ^{\prime \prime }\left(
r_{c}\right) -8\pi ^{2}\sigma _{c}^{2},  \label{29}
\end{equation}%
which upon considering (\ref{21}), it reduces to%
\begin{equation}
V^{\prime \prime }\left( r_{c}\right) =\psi ^{\prime \prime }\left(
r_{c}\right) -2\frac{\psi \left( r_{c}\right) }{r_{c}^{2}}.  \label{30}
\end{equation}%
Taking into account the above results in the equation of motion (\ref{24}),
up to the first nonzero term for the effective potential, the equation of
motion becomes%
\begin{equation}
\dot{a}^{2}+\frac{1}{2}V_{c}^{\prime \prime }\left( a-r_{c}\right)
^{2}\simeq \varepsilon ,  \label{31}
\end{equation}%
in which $V_{c}^{\prime \prime }=V^{\prime \prime }\left( r_{c}\right) $ and 
$\varepsilon >0$ is the initial small mechanical energy of the throat after
the perturbation. A derivative of (\ref{31}) with respect to the proper time
gives%
\begin{equation}
\ddot{a}+\frac{1}{2}V_{c}^{\prime \prime }a\simeq \frac{1}{2}V_{c}^{\prime
\prime }r_{c}.  \label{32}
\end{equation}%
The latter implies that with $V_{c}^{\prime \prime }>0$, the throat's radius
oscillates about its equilibrium radius i.e., $a=r_{c}$ and consequently the
TSW is stable against the radial linear perturbation. Otherwise, the
throat's radius expands or shrinks exponentially, and in either case, the
corresponding TSW is not stable. Therefore it is important to identify the
sign of $V_{c}^{\prime \prime }.$ To do so, once more we refer to the
properties of $r_{c}$ obtained from Eq. (\ref{16}). As we have mentioned
before, $r_{c}$ is the radius of the innermost null circular geodesics, and
as was proved in \cite{Hod2011}, the corresponding orbit possesses the
smallest period. Equation (\ref{16}) admits the critical point(s) for the
period function (\ref{14}) but to be the minimum period one should impose 
\begin{equation}
\left. T^{\prime \prime }\left( r\right) \right\vert _{r=r_{c}}>0.
\label{33}
\end{equation}%
Using the explicit expression of $T\left( r\right) $ in (\ref{14}), we find%
\begin{equation}
T^{\prime \prime }\left( r_{c}\right) =-\frac{2\pi r_{c}}{\sqrt{\psi \left(
r_{c}\right) }}\left( \frac{\psi ^{\prime \prime }\left( r_{c}\right) }{\psi
\left( r_{c}\right) }-\frac{2}{r_{c}^{2}}\right) ,  \label{34}
\end{equation}%
where we used $\psi ^{\prime }\left( r_{c}\right) =\frac{2}{r_{c}}\psi
\left( r_{c}\right) $ from (\ref{16}) or (\ref{R1}). From (\ref{34}) for the
period to be minimum at $r=r_{c}$ which is the radius of the innermost
circular null geodesics, we apply $T^{\prime \prime }\left( r_{c}\right) >0$
and consequently get the condition 
\begin{equation}
\left( \frac{\psi ^{\prime \prime }\left( r_{c}\right) }{\psi \left(
r_{c}\right) }-\frac{2}{r_{c}^{2}}\right) <0.  \label{35}
\end{equation}%
The latter \textbf{condition} directly implies 
\begin{equation}
V^{\prime \prime }\left( r_{c}\right) =\psi ^{\prime \prime }\left(
r_{c}\right) -2\frac{\psi \left( r_{c}\right) }{r_{c}^{2}}<0,  \label{36}
\end{equation}%
which completes our proof that the TSW with its throat at the innermost
photonsphere is unstable against a radial linear mechanical perturbation.
Before we finish this section it is remarkable to compare the equation of
motion of the throat i.e., Eq. (\ref{24}) with the circular motion of a null
particle outside the black hole (\ref{1}). As we have proved in a very
recent work \cite{Habib2024}, the motion of photons on the innermost
photonsphere is unstable (see also \cite{D1}). Furthermore, if Eq. (\ref{15}%
) yields more than one critical point since the smallest and the largest
radii correspond to the local minimum period due to the fact that $T\left(
r_{c}\rightarrow r_{+}\right) \rightarrow \infty $ and $T\left(
r_{c}\rightarrow \infty \right) \rightarrow \infty $, the number of critical
points is odd. In such cases, there is at least one local maximum for the
period and therefore $T^{\prime \prime }\left( r_{c}\right) <0$ and
consequently $V^{\prime \prime }\left( r_{c}\right) >0$ which implies the
corresponding TSW is stable. It should be added that there are special cases
where the two light rings can be degenerate that is characterized by $%
T^{\prime }\left( r_{c}\right) =T^{\prime \prime }\left( r_{c}\right) =0$\
which from (\ref{16}) and (\ref{34}) it becomes $r_{c}\psi ^{\prime \prime
}\left( r_{c}\right) =\psi ^{\prime }\left( r_{c}\right) .$

\section{Conclusion}

In this study, we have analytically proved that the instability of the STSW
supported by a barotropic surface fluid with the equilibrium throat at $%
a_{0}=3M$ is not a coincidence. In doing so, we started from a generic
static spherically symmetric black hole and proved that the corresponding
TSW constructed by applying the cut-and-paste method and powered by a
generic barotropic surface fluid is unstable when its radius is the same as
the innermost circular orbit of the black hole. This instability is also
connected to the instability of the innermost photonsphere of the black hole
as well (see \cite{D1,Habib2024}). It should be add that the instability
issue of the static spherically symmetric TSW supported by a barotropic
perfect surface fluid can be solved by considering a variable EoS for the
matter present on the throat. This is what Varela in \cite{Varela2015}
suggested for STSW. In this research, we studied and presented the
connection between the stability of the photon sphere and a thin-shell
wormhole in the static spherically symmetric black hole. The existence of a
similar connection in the rotational black hole is an open problem that we
shall address in a separate work.

\end{document}